# Dual-LED-based multichannel microscopy for whole-slide multiplane, multispectral, and phase imaging

*Jun Liao[1], Zhe Wang[1], Zibang Zhang[1,2], Zichao Bian[1], Kaikai Guo[1], Aparna Nambiar[1], Yutong Jiang[1], Shaowei Jiang[1], Jingang Zhong[2], Michael Choma[3], and Guoan Zheng[1,*]*

[1] Department of Biomedical Engineering, University of Connecticut, Storrs, CT, 06269, USA
[2] Department of Optoelectronic Engineering, Jinan University, Guangzhou 510632, China
[3] Department of Radiology & Biomedical Imaging, Biomedical Engineering, Applied Physics, and Pediatrics, Yale University, CT, 06520, USA

We report the development of a multichannel microscopy for whole-slide multiplane, multispectral, and phase imaging. We use trinocular heads to split the beam path into 6 independent channels and employ a camera array for parallel data acquisition, achieving a maximum data throughput of ~1 gigapixel per second. To perform single-frame rapid autofocusing, we place two near-infrared LEDs at the back focal plane of the condenser lens to illuminate the sample from two different incident angles. A hot mirror is used to direct the near-infrared light to an autofocusing camera. For multiplane whole slide imaging (WSI), we acquire 6 different focal planes of a thick specimen simultaneously. For multispectral WSI, we relay the 6 independent image planes to the same focal position and simultaneously acquire information at 6 spectral bands. For whole-slide phase imaging, we acquire images at 3 focal positions simultaneously and use the transport-of-intensity equation to recover the phase information. We also provide an open-source design to further increase the number of channels from 6 to 15. The reported platform provides a simple solution for multiplexed fluorescence imaging and multimodal WSI. Acquiring an instant focal stack without z-scanning may also enable fast 3D dynamic tracking of various biological samples.

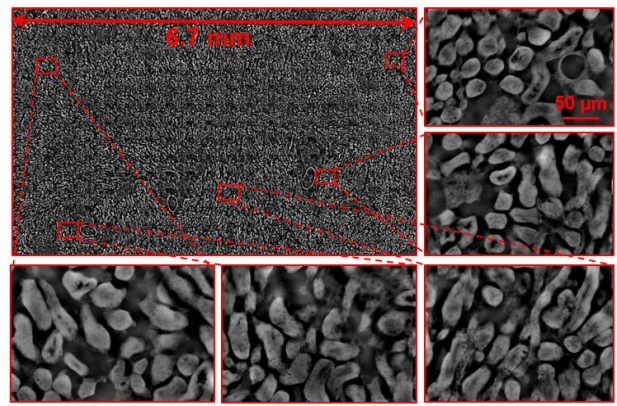

## 1. Introduction

Whole slide imaging (WSI) systems convert the conventional biological samples into digital images that can be analyzed with computers and shared through the internet[1]. It has undergone a period of rapid growth and expansion catalyzed by changes in imaging hardware and gains in computational processing. However, there are some technical challenges associated with the conventional WSI platform. First, conventional WSI acquires 2D images of thin sections. For thick sections such as cytology slides, a focal stack of the 3D cells and cell clusters is needed. Axial scanning via conventional WSI platform leads to a tradeoff between system speed and the number of focal planes. Second, immune-

---

histochemistry (IHC) has been used as an adjunctive tool to evaluate protein-expression patterns in tissue. This process assists in diagnosis by finding protein-expression patterns that correlate with the type and location of tumor[2]. One key consideration in IHC is to adopt multiplexed antibody staining to facilitate better quantitative studies. Multispectral imaging has been adopted for analyzing pathology slides stained with multiple antibodies[3]. Conventional multispectral system sequentially tunes the filter to different spectral bands and acquire the corresponding images, leading to a tradeoff between system speed and the acquired spectral channels. Third, it has been shown that the phase information (optical path length difference) of tissue sections is able to reveal the molecular scale organization of the sample. Whole-slide phase imaging may, therefore, enable label-free automatic tissue screening[4]. However, it is difficult to employ a regular microscope for whole-slide phase imaging. Fourth, in conventional WSI imaging systems, autofocusing is the most challenging issue to overcome and has been cited as the culprit for poor image quality in histologic diagnosis[5]. It is highly desired to develop a cost-effective plugin module for rapid autofocusing.

In this paper, we report the development of a multichannel microscope platform for whole-slide multiplane, multispectral, and phase imaging. Our implementation is built upon an existing regular microscope with straightforward modification. The contribution of this paper is threefold. First, we use commercially-available trinocular prisms to split the beam path into different independent channels. Each channel can be used to acquire sample information at one focal plane, one spectral band or one polarization state. We demonstrated a 6-channel WSI imaging platform using this strategy with minimum modification to an existing regular microscope. The achievable data throughput of the 6-channel platform exceeds 1 gigapixel per second and it allows for continues data streaming. We also provide an open-source design to further increase the independent channels from 6 to 15. Second, we show that we can use 3-channel multiplane data to recover the phase information using the transport of intensity equation (TIE). Since no axial scanning is needed, the reported approach may find applications in imaging fast-moving unstained biological sample such as cilia. Third, we provide an open-source optical design (separated from the multichannel hardware) for single-frame rapid autofocusing. In this design, we place two infrared LEDs at the back focal plane of the condenser lens to illuminate the sample from two different incident angles. A 45-degree hot mirror is placed at the objective-prism port to direct the infrared light to an autofocusing camera. The captured image from the autofocusing camera contains two copies of the sample separated by a certain distance. By identifying this distance, we can recover the defocus distance of the sample without z-scanning. In contrast to our previous single-frame autofocusing scheme[6], the proposed module uses two-angle illumination instead of two-pinhole modulation. It has all advantages of the previous design while requires little optical alignment and is ready for plug-and-play operation.

In the following, we will first report the 6-channels platform using the trinocular prisms. We will then report the autofocusing module using two near-infrared LEDs. Finally, we summarize the work and discuss the future directions.

## 2. Multichannel microscopy

The use of multiple cameras for parallel acquisition in microscopy has been demonstrated in multiplane microscopy with 2-4 cameras[7, 8]. Previous multiplane implementations, however, require the use of bulky optical relay to divide the beam path and have difficulties on expanding the independent channels beyond 4. It is also possible to use one camera and a diffraction grating to acquire information at multiplanes[9, 10]. However, a special dispersion compensation element is needed in this case and there is a trade-off between the field of view of one channel and the total number of channels.

Recently, a camera array has been employed in a light field microscope to acquire different perspectives of 3D samples[12, 13]. The acquired images are then used to perform 3D light field refocusing. However, even with the recent development of light field deconvolution [14, 15], there is still a resolution reduction compared to the diffraction limit of the employed objective lens. In many WSI applications such as digital pathology, achieving diffraction-limited resolution is of most importance to the users, and thus, light field microscopy may not be a good solution in this regard.

Different from the previous implementations, we employ an often-ignored component in a regular microscope -- the trinocular prism[16] for building our multichannel platform. In a regular microscope, the trinocular prism splits the light beam into 3 different channels, one for the camera port at the top and two others for the eyepieces, as shown in Fig. 1(a2)-1(a3). In Fig. 1(a3), we replaced the eyepiece tubes with a 3D-printed plastic attachment kit for housing the camera. Therefore, we can readily convert a regular microscope into a 3-channel microscope with minimum modification and without any additional component. These 3 independent channels can be used to image different focal planes, different spectral bands, and different polarization states. Such a simple implementation may enable the wide dissemination of the multi-channel microscopy for a wide range of applications in biological and clinical labs, including multiplexed fluorescence imaging, super-resolution temporal imaging[17, 18], 3D localization based super-resolution imaging[19], among others.

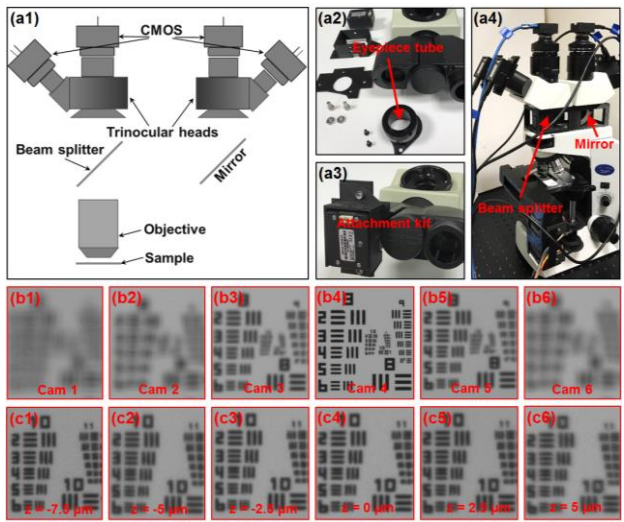

**Figure 1** (a1) Schematics of multichannel microscopy. (a2)-(a3) We replace the eyepiece tube with a custom 3D-printed camera attachment kit. (a4) We employ two trinocular heads with 6 cameras in the prototype setup (Olympus CX 41). (b) The captured 6 images of the USAF target. (c) Resolution characterization by placing the resolution target at the corresponding focal planes. A detailed instruction can be found at our open-source protocol[11].

To further increase the number of independent channels, we use two trinocular heads in our prototype setup in Fig. 1(a4) and each head provides 3 channels. The relative optical power for these 6 ports are 1, 0.5, 0.5, 1, 0.5, and 0.5 ('1' for the camera ports and '0.5' for the eyepiece ports). To select different focal planes for different cameras, we added spacers to change the distance between the cameras and the tube lens. For the eyepiece ports, we used flat washers as spacers for coarse adjustment and tapes for fine adjustment. For one camera port, we used a 5 mm c-mount extension ring and tapes as the spacer. We used a resolution target and calculated the Brenner gradient value to calibrate the location of the focal planes (the precision is less than 0.3-micron depth of field). A small z-translator would make the focal plane positioning more flexible and convenient. The focal planes for the 6 channels are at z = -7.5 µm, -5 µm, -2.5 µm, 0 µm, -2.5 µm, and 5 µm for a 20X, 0.75 NA objective lens. Figure 1(b1)-(b6) show the captured images of a USAF resolution target from the 6 channels (monochromatic camera: CM3-U3-50S5M-CS, PointGrey, 5 megapixels at 35 fps). To characterize the imaging performance, we moved the resolution target to different focal planes and captured the images using the corresponding channels in Fig. 1(c). All 6 channels can resolve group 10, element 6 of the resolution target (0.275 µm linewidth). We do not observe resolution loss by adding the spacer to the camera attachment kit.

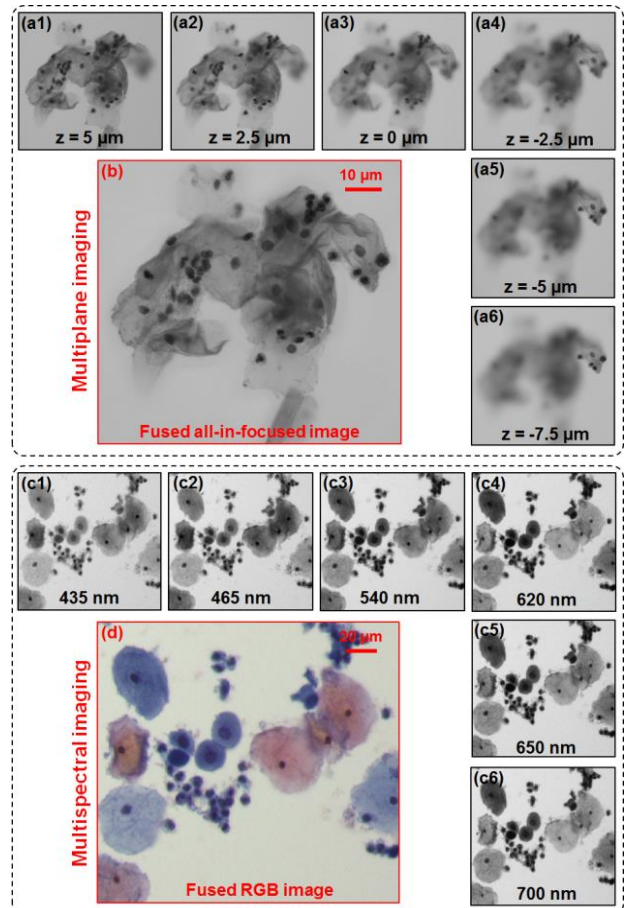

**Figure 2** Multiplane and multispectral imaging using the 6-channel prototype. (a1)-(a6) Multiplane images using the 6 channels. (b) The fused all-in-focus image using (a1)-(a6). (c) Multispectral images using the 6 channels. (d) The fused color image using the R, G, and B channels.

In Fig. 2(a), we use the 6-plane platform to acquire images of a Pap smear sample. We can see that different parts of the samples are in focus at different channels. In Fig. 2(b), we fuse all 6-plane images together to extend the depth of field and all regions of the sample are in focus in this case[20]. Similarly, the 6 independent channels can be used for multispectral imaging. In Fig. 2(c), we remove the spacers of the attachment kits so that all cameras have the same focal plane. We then add 6 bandpass filters in front of the cameras and capture the corresponding images. The central wavelengths of the bandpass filters are 435, 465, 540, 620, 650 and 700 nm, with a ~80 nm bandwidth. Figure 2(d) shows combined color image using the 465-, 540-, and 620-nm channels. The pixel throughput of the 6-channel prototype platform exceeds 1 gigapixel per second (each channel captures 5-megapixel images at 35 fps). Figure 3 and Visualization 1 show the multiplane video of a living Daphnia sample (Caralina Biological Inc.). The capability of recording multiplane information without z scanning may find important applications in 3D fast dynamics tracking.

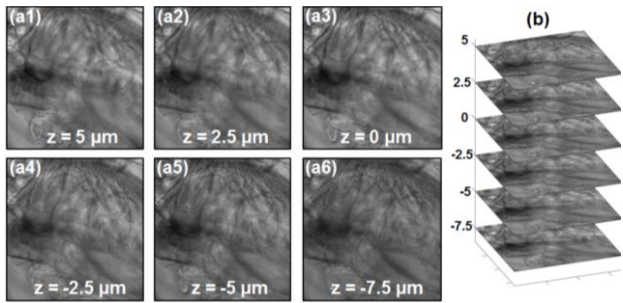

**Figure 3** (Visualization 1) Multiplane microscopy for tracking 3D dynamic of a living Daphnia sample. We used hardware triggering signals to synchronize the 6 cameras. (a) The multiplane images. (b) The focal stack of the 6 planes.

Another application for the multichannel microscopy is to recover the phase information of transparent samples. In Fig. 4(a), we acquired 3 images of an unstained mouse kidney slide (Molecular Expressions Inc.) at 3 different focal positions at the same time. We then used the transport of intensity equation (TIE)[21-23] to recover the phase image of the sample, as shown in Fig. 4(b). The TIE describes the relationship between the intensity and phase distribution while the wave is propagating along the axial direction. It is first put by Teague in Ref. [23], which showed that the phase can be determined by measuring intensity images at different focal planes. In our implementation, we used an open-source fast-Fourier-transform-based TIE solver[24] (http://www.scilaboratory.com/h-col-123.html) to recover the phase image. The accuracy of this method has been validated using microlens array[25]. Since we can record multiplane information in high speed, the reported approach may be able to recover the phase images of fast-moving samples such as cilia in a post-acquisition processing manner.

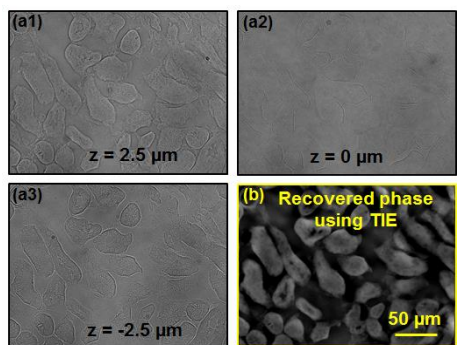

**Figure 4** Multiplane microscopy for recovering the phase information of an unstained mouse kidney section. (a1)-(a3) The three images captured at z = 2.5 µm, 0 µm, and -2.5 µm positions. (b) The recovered phase image using TIE. It took ~0.7 seconds to recover a 1520-by-1520-pixel image using an Intel i5 CPU.

In many biomedical experiments, one needs to capture fluorescence images at different emission bands. The reported platform can simultaneously capture multiband fluorescence images without mechanically switching the filter cube. This may be useful for tracking fast-moving biological samples with multiband fluorescence signals. As shown in Fig. 5(a), we used 3 cameras for image acquisition and a mouse kidney slide as the sample (stained with Alexa 568, Alexa 488, and DAPI, Molecular Expressions Inc.). In the epi-illumination arm, we used a standard DAPI filter cube to generate excitation light (central wavelength: 360 nm). In the detection path, we placed three different emission filters in front of the three cameras and their corresponding images are shown in Fig. 5(b1)-5(b3). Figure 5(c) shows the combined image of the three fluorescence channels (integration time: ~0.1 seconds). One concern for this setup is that fluorescence light is weaker due to the beam splitting. We argue that the exposure time can be on proportionally longer as no filter switching is needed. If the spectral bands are not equally bright, we can still perform synchronous imaging between different channels. In this case, the exposure time will be set by the brightest channel. Post-acquisition averaging can be used to increase the SNR of the dim channels if needed (assuming read noise is low).

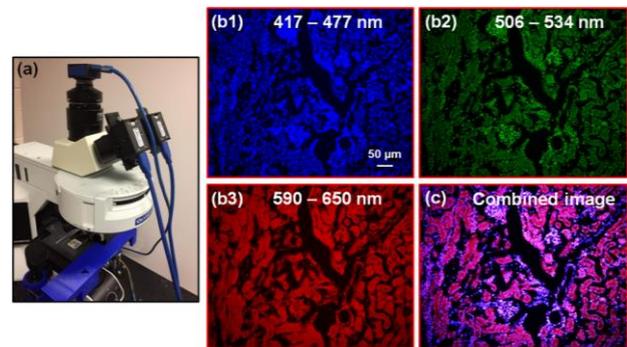

**Figure 5** Multiplexed fluorescence imaging without switching the filter cube. (a) We attached 3 cameras to a regular epi-illuminated fluorescence microscope (Olympus BX 43 with a 20X 0.75 NA objective lens). (b) The captured multiband fluorescence images of the sample. (c) The combined multiband fluorescence image.

## 3. Whole slide imaging with a dual-LED autofocusing module

One key consideration in WSI is to perform autofocusing in high speed. Recently, we have demonstrated the use of a two-pinhole modulated camera for single frame rapid autofocusing[6]. In that platform, the two-pinhole aperture is placed at the Fourier plane of the imaging system. Here, instead of using the two-pinhole modulation scheme, we place two 740-nm LEDs (1516-1213-1-ND, Digikey) at the back focal plane of the condenser lens for sample illumination (Fig. 6(a1)). These two LEDs illuminate the sample from two different incident angles and can be treated as spatially coherent light sources.

If the sample is placed at an out-of-focus position, the captured image will contain two copies of the sample separated by a certain distance. By identifying this separation through the autocorrelation function, we can directly recover the defocus distance without z-scanning. As shown in Fig. 6(a1) and 6(a2), we have designed an add-on kit that attaches to the polarization port of the microscope platform. This kit contains a 45-degree hot mirror (43-955, Edmund Optics) and a CCTV lens (SainSonic 50mm f/1.4, Amazon) to direct the infrared light to the camera. Figure 6(a2) shows the entire multi-channel WSI platform with the autofocusing add-on kit. Figure 6(b1) shows a raw image captured by the camera and Fig. 6(b2) shows its autocorrelation function by which we can identify the separation distance $x_0$. Figure 6(c1)-6(c3) show three captured images at different focal planes. Figure 6(c4) shows the measured relationship between separation distance $x_0$ (in pixel) and the defocus distance (in µm).

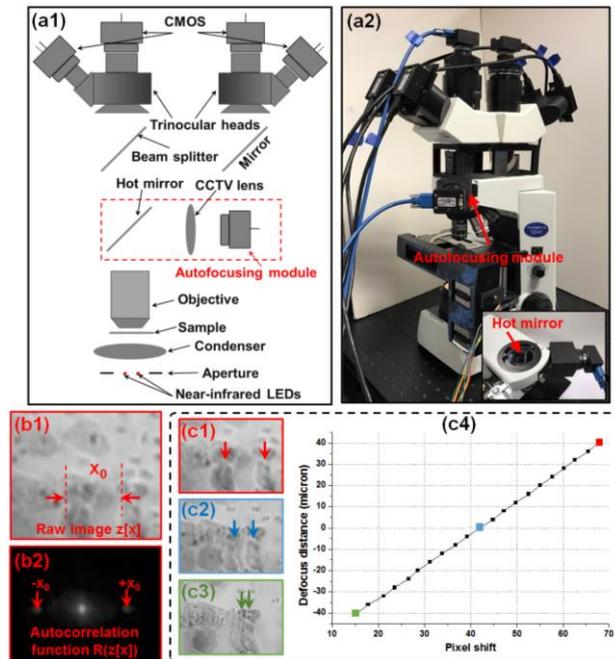

**Figure 6** Multiplane and multispectral WSI using a dual-LED autofocusing module. (a1) Two infrared LEDs are placed at the back focal plane of the condenser lens. (a2) The auto-focusing add-on kit is inserted into the polarization port of an Olympus CX 41 microscope. (b) The captured raw image from the autofocusing module and its autocorrelation function. (c) The relationship between the separation $x_0$ and the defocus distance.

Based on the autofocusing add-on kit in Fig. 6, we can perform WSI using the multichannel microscope platform. In Fig 7(a)-(b), we show the captured whole-slide multiplane and multispectral images of a Pap smear sample. For the multiplane WSI, we acquired 6 images at different focal positions at the same time. For the multispectral WSI, we acquired 6 images of the same focal plane but with 6 different spectral bands at the same time. Similarly, we can also perform whole slide phase imaging as shown in Fig. 8. In this case, we used 3 channels to simultaneously acquire images at z = -2.5 µm, 0 µm, and +2.5 µm. We then recovered the phase images and stitched them to form a whole slide imaging in Fig. 8.

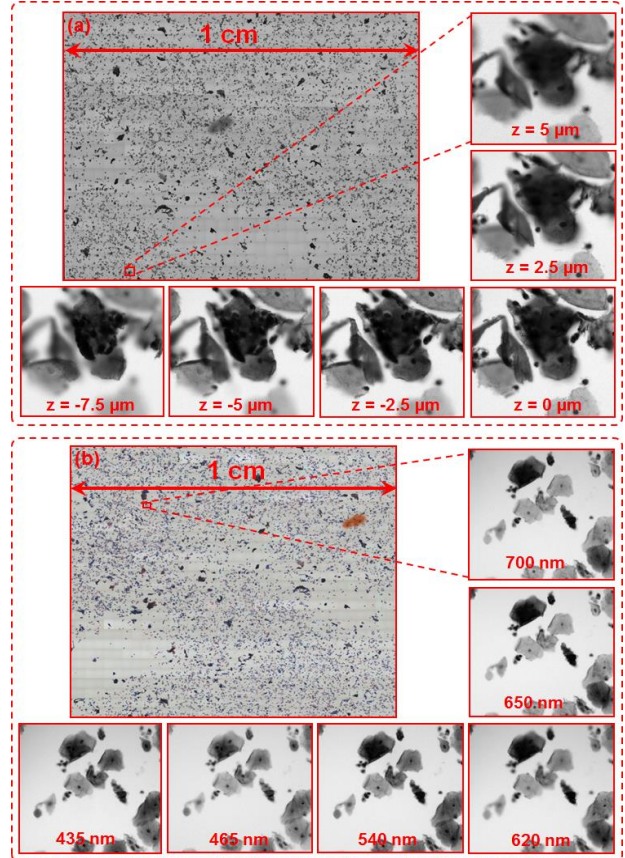

**Figure 7** Multiplane (a) and multispectral (b) WSI using a dual-LED autofocusing module.

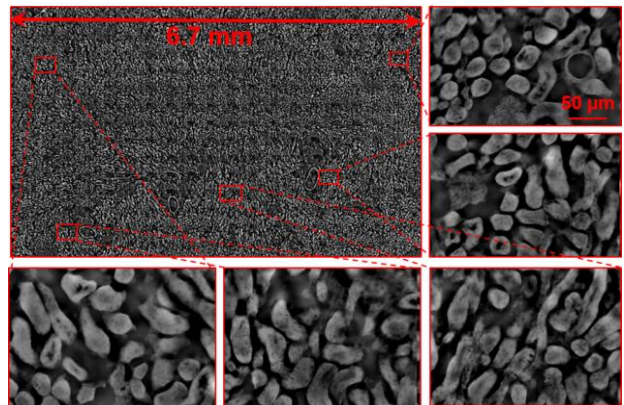

**Figure 8** Whole slide phase image of an unstained mouse kidney sample. We used 3 channels to simultaneously acquire images at z = -2.5 µm, 0 µm, and +2.5 µm. The recovered phase images were then stitched to form the whole slide image. The full whole slide image can be found: http://gigapan.com/gigapans/195918.

Compared to the two-pinhole modulation scheme, there are several advantages of the reported dual-LED

autofocusing module. First, no pinhole aperture is needed at the Fourier plane and the captured image contains all the high-resolution details. We do not need to relay the Fourier plane out of the objective lens and no aperture alignment is needed. Second, the use of infrared light will not affect the visible band and all visible photons remain unchanged at the detection path. Third, the use of polarization port allows a simple plug-and-play operation. There is no modification to the epi-illumination arm of conventional fluorescence microscope platform. Fourth, the position of the pupil plane may change when we switch to a different objective lens. Since we do not use aperture at the pupil plane, it can be used for autofocusing regardless the employed objective lens. Lastly, we provide all 3D design files of this module[11].

## 4. Conclusion

In summary, we have reported the development of a multichannel microscope platform for whole-slide multiplane, multispectral, and phase imaging. The contribution of this paper is threefold. First, we demonstrate the use of trinocular prisms to divide the beam path to multiple independent channels. By using this strategy, we can readily convert a regular microscope into a 3-channel microscope with minimum modification and without any additional component. These 3 independent channels can be used to capture images at different focal planes, at different spectral bands, with different polarization states, and with different exposure times. Such a simple implementation may enable the wide dissemination of the multichannel microscopy for a wide range of applications in biological and clinical labs. Second, we show that 3-channel multiplane data can be used to recover the phase information using the TIE. Since no axial scanning is needed, the reported approach may find applications in imaging fast-moving unstained biological sample such as cilia. Third, we report a dual-LED autofocusing module that can be directly inserted into the polarization port for single-frame rapid autofocusing. No pinhole modulation is needed in the reported module, allowing a simple plug-and-play operation without precise optical alignment.

To the best of our knowledge, there is no previous report on using the eyepiece ports for high-resolution microscopy. There may be two reasons for this. First, the aberration-corrected image plane is within the eyepiece tube and one needs to remove the eyepiece tube to access it. If the image sensor is placed outside the eyepiece tube, spherical aberrations will be introduced to the captured image. Second, the commercially available 1X or 0.5X eyepiece adapter will introduce significant field-dependent aberrations for high NA objective lenses.

The development of the reported platform is timely as well. Driven by cellphone camera market, the performance of cost-effective CMOS camera has been substantially improved in the past few years. The dark noise of the cost-effective image sensor in our platform (Sony IMX264) is 2.29 electrons and the dynamic range is 70.97 dB, which are comparable to many high-end CCDs or scientific CMOS cameras. The reported platform has 6 independent channels. We can further increase the 6 channels into 15 channels by adding 3 more trinocular heads using both the polarization port and the epi-illumination arm[11]. If the side port is available, it can also be used to increase the number of output channels.

One of our on-going efforts is to explore the use the reported platform for super-resolution temporal imaging. By triggering the cameras at slightly different times, we can achieve an imaging frame rate (throughput) that is one order of magnitude higher than that of current camera bandwidth. This will be a simple and effective approach to study the fast dynamic of biological samples. Finally, we have made all 3D-printing design files and protocol open source. Interested readers can download them through[11].

**Acknowledgments** This work was in part supported by NSF DBI 1555986, NIH R21EB022378, and NIH R03EB022144. K. Guo acknowledges the support of FEI fellowship by the FEI Company. Z. Zhang acknowledges the support of the China Scholarship Council. M. A. Choma was supported by NIH1R01-HL118419 and R21 EB017935-01A1.

## References

[1] L. Pantanowitz, J. H. Sinard, W. H. Henricks, L. A. Fatheree, A. B. Carter, L. Contis, B. A. Beckwith, A. J. Evans, A. Lal, A. V. Parwani *Archives of Pathology and Laboratory Medicine*. **2013**, *137*, 1710-1722.
[2] F. Ghaznavi, A. Evans, A. Madabhushi, M. Feldman *Annual Review of Pathology: Mechanisms of Disease*. **2013**, *8*, 331-359.
[3] R. M. Levenson, J. R. Mansfield *Cytometry part A*. **2006**, *69*, 748-758.
[4] Z. Wang, K. Tangella, A. Balla, G. Popescu *Journal of biomedical optics*. **2011**, *16*, 116017-1160177.
[5] J. R. Gilbertson, J. Ho, L. Anthony, D. M. Jukic, Y. Yagi, A. V. Parwani *BMC clinical pathology*. **2006**, *6*, 1.
[6] J. Liao, L. Bian, Z. Bian, Z. Zhang, C. Patel, K. Hoshino, Y. C. Eldar, G. Zheng *Biomedical Optics Express*. **2016**, *7*, 4763-4768.
[7] P. Prabhat, S. Ram, E. S. Ward, R. J. Ober *IEEE transactions on nanobioscience*. **2004**, *3*, 237-242.
[8] S. Ram, P. Prabhat, J. Chao, E. Sally Ward, R. J. Ober *Biophysical Journal*, *95*, 6025-6043.

[9] S. Abrahamsson, J. Chen, B. Hajj, S. Stallinga, A. Y. Katsov, J. Wisniewski, G. Mizuguchi, P. Soule, F. Mueller, C. D. Darzacq, X. Darzacq, C. Wu, C. I. Bargmann, D. A. Agard, M. Dahan, M. G. L. Gustafsson *Nat Meth*. **2013**, *10*, 60-63.

[10] S. Abrahamsson, R. Ilic, J. Wisniewski, B. Mehl, L. Yu, L. Chen, M. Davanco, L. Oudjedi, J.-B. Fiche, B. Hajj, X. Jin, J. Pulupa, C. Cho, M. Mir, M. El Beheiry, X. Darzacq, M. Nollmann, M. Dahan, C. Wu, T. Lionnet, J. A. Liddle, C. I. Bargmann *Biomedical Optics Express*. **2016**, *7*, 855-869.

[11] Open source protocol for the reported platform: https://figshare.com/s/a0d09c4bedf0b1ac8132

[12] X. Lin, J. Wu, G. Zheng, Q. Dai *Biomedical Optics Express*. **2015**, *6*, 3179-3189.

[13] J. Wu, B. Xiong, X. Lin, J. He, J. Suo, Q. Dai *Scientific reports*. **2016**, *6*.

[14] R. Prevedel, Y.-G. Yoon, M. Hoffmann, N. Pak, G. Wetzstein, S. Kato, T. Schrodel, R. Raskar, M. Zimmer, E. S. Boyden, A. Vaziri *Nat Meth*. **2014**, *11*, 727-730.

[15] M. Broxton, L. Grosenick, S. Yang, N. Cohen, A. Andalman, K. Deisseroth, M. Levoy *Optics Express*. **2013**, *21*, 25418-25439.

[16] K. Guo, J. Liao, Z. Bian, X. Heng, G. Zheng *Biomed. Opt. Express*. **2015**, *6*, 3210-3216.

[17] G. Bub, M. Tecza, M. Helmes, P. Lee, P. Kohl *Nature Methods*. **2010**, *7*, 209-211.

[18] A. Agrawal, M. Gupta, A. Veeraraghavan, S. G. Narasimhan in Optimal coded sampling for temporal super-resolution, Vol. (Ed.^Eds.: Editor), IEEE, City, **2010**, pp.599-606.

[19] S. R. P. Pavani, M. A. Thompson, J. S. Biteen, S. J. Lord, N. Liu, R. J. Twieg, R. Piestun, W. Moerner *Proceedings of the National Academy of Sciences*. **2009**, *106*, 2995-2999.

[20] S. Pertuz, D. Puig, M. A. Garcia, A. Fusiello *IEEE Transactions on Image Processing*. **2013**, *22*, 1242-1251.

[21] T. Gureyev, A. Roberts, K. Nugent *JOSA A*. **1995**, *12*, 1942-1946.

[22] N. Streibl *Optics communications*. **1984**, *49*, 6-10.

[23] M. R. Teague *JOSA*. **1983**, *73*, 1434-1441.

[24] C. Zuo, Q. Chen, A. Asundi *Optics Express*. **2014**, *22*, 9220-9244.

[25] C. Zuo, Q. Chen, H. Li, W. Qu, A. Asundi *Optics Express*. **2014**, *22*, 18310-18324.